# Design principles for the energy level tuning in donor/acceptor conjugated polymers

*Daniel Hashemi[‡], Xiao Ma[‡], Jinsang Kim and John Kieffer[*]*

Department of Materials Science and Engineering, University of Michigan, Ann Arbor, MI 48109

*Corresponding author.

[‡] These authors contributed equally to this work.

E-mail: kieffer@umich.edu;

Phone: 734-763-2595;

Fax: 734-763-4788.

ABSTRACT

To identify reliable molecular design principles for energy level tuning in donor/acceptor conjugated polymers (CPs), we studied the governing factors by means of *ab initio* calculations based on density-functional theory (DFT). We investigated a series of CPs in which we independently and systematically varied the electron withdrawing power of the acceptor unit and the electron donating power of the donor unit, while maintaining the same conjugated chain conformation. We observed that the introduction of a stronger acceptor unit, while keeping the same donor unit in the CP, lowers the LUMO level, but leaves the HOMO level almost unchanged. Conversely, enhancing the strength of the donor unit for the same acceptor unit raises the HOMO level, while maintaining the LUMO level. We identified strong correlations between the frontier orbital energy levels and the degree of orbital localization, which depends on the electron donating or withdrawing power of the molecular groups carrying the orbitals. Moreover, the HOMO/LUMO gap of the CP is directly proportional to the charge transfer between donating and accepting units, which provides a robust design criterion for CPs.

# 1 Introduction

Conjugated polymers (CPs) are widely used in organic photovoltaic (OPV) devices. The power conversion efficiency of these systems has exceeded 10 %.[1] However, the optical energy gaps of conventional organic materials are normally greater than 2 eV, covering only the visible range of the solar spectrum (300-650 nm).[2,3] Smaller energy gap organic materials are in demand to extend absorption to the long wavelength region to take advantage of the higher flux, red light and even near infrared region of the solar spectrum. Therefore, it has been a major effort to design and synthesize novel conjugated polymers with narrow energy gaps in recent years. However, the magnitude of the energy gap is not the sole criterion for molecular design.

The absolute values of the energy levels for the highest occupied (HOMO) and lowest unoccupied molecular orbitals (LUMO) must be compatible with those of the adjacent buffer or electrode layers. There are three principal ways to modify the frontier orbital energies: (i) enlarging of the π orbital systems;[4] (ii) incorporating planar fused aromatic ring systems such as quinoidal structures;[5] and (iii) incorporating alternating donor/acceptor functional units.[5,6] While these methods are very effective for reducing energy gaps, both options (i) and (ii) have the adverse effect of increasing HOMO energy levels, resulting in a decreased device open circuit voltage ($V_{oc}$). Method (iii) overcomes this problem by integrating electron-rich and electron-deficient molecular groups into the donor polymer layer, and thereby providing a local acceptor character that affects frontier orbital energy levels and possibly also exciton lifetimes. This combined donor-acceptor (D-A) strategy is now widely used to design efficient polymer photovoltaic materials.

In such copolymer systems, interactions between alternating D-A units are strong enough to cause the HOMO/LUMO gap to shrink.[7] This is associated with the reduction of bond-length alternation by enhancing the double-bond character between the repeat units of the CP[8] since they can accommodate the charges that are associated with mesomerism (D+A→D$^+$+A$^-$).[9]

Recently, Kim *et al*. have systematically explored a series of CPs by varying the acceptor units while keeping the donor unit unchanged. [10] It was shown that a more electronegative acceptor unit lowers both the HOMO and the LUMO energies of the CP. However, the drop in the LUMO energy is significantly more pronounced than that in the HOMO energy, resulting in a narrowing of the HOMO/LUMO gap. The nearly unchanged HOMO energy suggests energy level changes are predominantly controlled by the nature of the donor unit. The phenomenon of HOMO/LUMO gap narrowing due to acceptor unit variation was also verified by the blue shift in the absorption spectrum reported in the same study.

In the present investigation, first-principles calculations are used to systematically investigate the aforementioned phenomena and to elucidate design principles for conjugated polymer systems. By methodically scrutinizing several key factors that influence the frontier orbital energy levels and the HOMO/LUMO gap, important principles to design D–A copolymers are revealed. This paper is organized as follows: in Section 2, a brief description of the theory and computational details is given. In section 3, the acceptor unit is varied in four D-A polymers with the same donor unit as (Figure 1), previously synthesized by Kim et al.,[10]. Molecular structures, HOMO-LUMO levels and HOMO/LUMO gaps of the isolated donors, the isolated acceptors, the D-A oligomer and the synthesized D-A polymers, as well as the role of charge localization and intramolecular charge transfer are discussed in detail. After determining consistency in our simulation results and the experimental findings for the known systems, we expand our study to include newly conceived CPs with varying acceptor units while the donor unit is kept constant. This allows us to further validate the mechanisms responsible for the observed behaviors, and identify robust molecular design criteria for energy level tuning in CPs. Finally, we summarize and conclude in Section 4.

## 2  Computational details

The electronic structure and properties of the conjugated polymers were investigated using density functional theory (DFT) calculations, carried out with Gaussian09,[11] employing the hybrid functional B3LYP and fully periodic boundary conditions (PBC). After geometry optimization, electronic structure calculations were performed using a larger 6-311G(d,p) basis set with 5 k-points along the one dimensional Brillouin zone. The HOMO/LUMO gap of a polymer is calculated as the minimum difference between the HOMO and the LUMO energy levels at a given k-point. The energy levels of the HOMO and LUMO are calculated as the maximum of the HOMO and the minimum of the LUMO, respectively. The Mulliken[12] charge distribution on each atom is also obtained after geometry optimization.

## 3  Results and discussion

### 3.1  Varying the acceptor in conjugated oligomers or polymer for a given donor

The degree of planarity within the conjugated backbone not only influences the electronic and optical properties considerably, but from a practical computational viewpoint, it can also affect the cost associated with the calculations and analysis of longer oligomers.[13] Therefore, CP backbones are constructed from thiophene derivatives to ensure a rigid-planar conformation. Figure 1 shows all the polymers that have been synthesized by Kim et al.,[10] i.e. PBTCT, PBTCTO, PBTPDO, and PBTFDO. They all have the exact same zig-zag shaped backbone geometry. Among them, PBTCT and PBTCTO exhibit twist angles between the D-A repeating units of 7.5° and 7.9°, respectively, which are slightly larger than the twist angles for PBTPDO and PBTFDO (≈1°–3°). An additional geometric consideration is the orientation of the individual components along the conjugated backbone.

D-A components are able to stabilize particular configurations since they often contain heteroatoms, that introduce intramolecular van der Waals and other electrostatic interactions.[14,15] For testing purposes, we studied the bithienyl oligomer, which consists of two adjacent thiophene units. Depending on the positioning the thiophene units, we can have the *cis* or *trans* conformation, as shown in Figure 2. According to our calculations, the *trans* is energetically more stable by 10 meV than the *cis* conformation in the gas phase. This is because the lone pairs on the sulfur atoms repel each other in the *cis* conformation. Consequently, we built the backbone of the polymers with a *trans* conformation initially and later optimized it using DFT calculations.

The hybridization of molecular orbitals as a result of combining different monomeric groups may lead to unexpected new energy levels. This is because the HOMO and LUMO energies have a strong dependence on the degree of electron delocalization within the CP's effective conjugation length,[16] which is affected by the conformation of the backbone.[17,18] Therefore, the prediction of the HOMO/LUMO gap of a CP based on the respective HOMO/LUMO energies of the donor unit or acceptor unit is not a trivial task.

To identify the governing principles more clearly, we begin by comparing the frontier orbital energy levels of the simplest oligomer, a single D-A repeat unit. Figure 3 shows the structures of the donor unit and four different acceptor units. We also included a donor-donor (D-D) oligomer and acceptor-acceptor (A-A) oligomer for comparison. As shown in Figure 4, our calculations reveal that upon combining a donor and an acceptor unit into a single D-A repeat unit, the resulting HOMO level mirrors that of the isolated donor unit and the LUMO level to that of the isolated acceptor unit, in almost all cases. The only exception is seen in the PBTCT monomer due to the comparable electron donating strength of the donor and acceptor in this molecule.

Bonding donor and acceptor groups to each other causes some amount of charge to transfer between the two, as illustrated in Figure 5. The strength of the electron donating or withdrawing behavior can be measured by the excess of charge compared to the isolated units.[19] For PBTCT the electron donating strength of the donor unit is only slightly larger than that of the acceptor unit. In such a CP, neither the donor nor the acceptor strongly influence the CP's electron donating or withdrawing characteristics. Also, unlike the other D-A monomers, there is a 20-30% uncertainty in the calculated HOMO/LUMO gap. The use of weak donor and acceptor units does not yield low HOMO/LUMO gap materials.[20] Moreover, the D-D oligomer and A-A oligomer both show a narrowed HOMO/LUMO gap as compared to the isolated donor and acceptor units. We therefore conclude that PBTCT represents an atypical outlier in our analysis, and that the energy levels of a D-A monomer can be well predicted using the HOMO level of the donor unit and the LUMO level of the acceptor unit, provided that the two units have a sufficiently large difference in their electron donating/withdrawing strength.

Despite the success of predicting the HOMO and LUMO levels of isolated D-A groups from the characteristics of the individual units, there is not satisfactory agreement with experimental observations; in experimental systems CPs consist of a large number of repeat units, and the extended conjugation of orbitals is a hallmark of these polymers that is hard to reproduce in simulations. To elucidate the effect of conjugation length on electronic properties, we gradually increase the number of repeat units for our calculations, from 1 to 4, which is our computational limit. However, by constructing a simulation cell that is periodic in the direction of the polymer backbone we can mimic the case of a CP with continuous conjugation, or a quasi-infinite number of repeat units. While periodic boundary conditions do not allow for exploring the extent of conjugation lengths applicable to real CPs, it eliminates symmetry breaking end groups and provides a more accurate estimation of a central segment of the polymer. The periodic unit cell in this case contains two repeat units.

Figure 6 shows the HOMO and LUMO levels calculated for these CP oligomer configurations as well as the energy levels of experimentally synthesized CPs as measured by cyclic voltammetry (CV).[10] Our calculations demonstrate that there is a minimal decrease in HOMO energies for all CPs as we change the number of repeat units, but the LUMO energies decrease more dramatically. Furthermore, more electronegative acceptor units cause a more drastic decrease in the LUMO energy. The lowered LUMO energies leads to a narrowing of the HOMO/LUMO gap in the sequence from PBTCT to PBTFDO, as we can see from the graphs in Figure 6(b) and the data in

Table 1. The observed trend agrees well with the experimental findings. We found that the LUMO levels of periodic unit cells in our calculation are consistently higher than the experimentally measured ones, which is most likely due to the fact that even after eliminating termination groups via periodic boundary conditions, our simulations cannot reach the effective conjugation lengths that apparently govern the behavior of experimental systems. Knowing this, our calculations support the expectation that larger conjugation lengths lower the LUMO energy. Zhang and Musgrave[21] showed that even though the hybrid DFT can predict the HOMO/LUMO gap with relative accuracy, the predicted HOMO and LUMO energies are inaccurate.

In order to elucidate the role of conjugation length on the energy levels of CPs, their frontier orbitals are shown in Figure 7. We compare our findings for tetramers and quasi-infinite periodic chains. For CP tetramers, the HOMO orbitals are generally less localized than the LUMO orbitals, but no clear correlation between the spatial arrangement of frontier orbitals and the HOMO-LUMO gap magnitude is apparent. Moreover, depending on the molecule, the HOMO or the LUMO orbitals shift to one side of the tetramer, effectively breaking orbital symmetry. After verifying that neither symmetric nor asymmetric termination of the tetramer can eliminate this incongruous behavior we attribute it to a numerical artifact. Indeed, upon removing terminating molecular groups altogether by connecting the free ends into a periodic chain, this erratic behavior vanishes.

After this correction, we see that the HOMO orbitals remain uniformly delocalized while the LUMO orbitals increasingly localize on acceptor units as the elements in this moiety become more electronegative. Taking PBTCT and PBTFDO as the opposite extremes of this behavior, we see that a fair amount of LUMO density of PBTCT polymer can still be found within the thiophene moieties of the donor units, the LUMO density of PBTFDO polymers is completely concentrated on the acceptor units. A similar trend has been observed by others: as the chain length of D-A polymers increases, the LUMO energy level decreases more rapidly than the HOMO energy level, producing a narrowed HOMO/LUMO gap due to changes in molecular orbital hybridization of the donor and acceptor units.[13] Our finding indicates that (i) the donor unit determines the HOMO levels of the CPs, (ii) the characteristics of the acceptor group allows one to estimate an approximate value for the LUMO level, (iii) the more the LUMO is localized, the lower the LUMO energy will be.

As already alluded to earlier, the amount of intramolecular charge transfer between donor and acceptor appears to be a another indicator for the extent to which the HOMO/LUMO gap is narrowed in CPs.[22] In particular, since all our CPs have the same donor, it is straightforward to evaluate the relative electron withdrawing strengths for the different acceptor units by simply comparing the charge differences between donor and acceptor upon combining these units using Mulliken charge analysis. It should be noted that we only compared the sum of the non-hydrogenized part of the donor unit, as the π-electrons on the conjugation backbone are dominant in intramolecular charge transfer.[23]

Figure 8 shows the total charge difference between donor and acceptor units as a result of the intramolecular charge transfer. A positive value indicates a departure of electrons from the donor. The CP monomer is omitted because we only examined central donor units to minimize any size effects. As expected, intramolecular charge transfer increases as the electronegativity of the acceptor unit increases. Furthermore, a rough proportionality (with a correlation coefficient of $R^2 = 0.78$) between the amount of charge transfer and the narrowing of the HOMO/LUMO gap is observed. We also notice that the amount of intramolecular charge transfer in PBTCT strongly depends on the position of the donor unit within the oligomer. This is reflected in the relatively large variation in the total charge in the donor units at different positions, particularly in the tetramer. Conversely, in the stronger D-A oligomers, e.g. PBTCTO, PBTPDO and PBTFDO, the amount of intramolecular charge transfer is unchanging, irrespective of the donor-acceptor pair position along the extent of the oligomer. Interestingly, the stronger the localization of the LUMO orbitals on the acceptor, the less variation is observed in the charge transfer as a function of the position along the oligomer.

**3.2 Varying the donor in conjugated oligomers or polymer for a given acceptor**

To further substantiate the trends observed when we vary the acceptor molecules, we investigated whether similar governing principles apply when altering the chemistry of the donor groups. To this end we conceived a series of systematic functionalizations and elemental substitutions within the donor unit of PBTFDO, creating molecular designs that allow us to control the HOMO level while pinning the LUMO level. As before, a major constraining factor is the maintenance of planar polymer conformations. After exploring several chemistries, the best design strategy turned out to be elemental substitution in the PBTFDO compound. Keeping the same acceptor atom, we replaced up to two sulfur atoms in the donor unit with Se or O to obtain a new donor unit, labeled as PBTFDO(Se), PBTFDO(2Se), PBTFDO(O), PBTFDO(2O), as is shown in Figure 9.

As demonstrated in literature, the energy gap of a furan-based CP is smaller than that of a thiophene-based CP with a similar structure.[24,25] This behavior is attributed to two factors: First, in comparison with thiophene, the five-membered ring of furan has weaker steric hindrance to adjacent units, because of a smaller diameter of the oxygen atom. Therefore a planar structure and a well conjugated backbone is formed. [26] Second, the delocalization is not as extensive in furan because of the high electronegativity of oxygen, so the lone pair is held more tightly by the oxygen. Therefore, aromatic stabilization is weak in furan, which enhances its electron donor activity. [27] Based on these findings, we would expect a narrowed HOMO/LUMO gap for oxygen substitution, and a broadened HOMO/LUMO gap for selenium substitution.

All of the newly designed CPs have a zig-zag conformation similar to that of PBTFDO. The twist angles between the D-A repeating units are 1.6°, 1.1°, 16.1°, 12.5° for PBTFDO(2Se), PBTFDO(Se), PBTFDO(O), PBTFDO(2O), respectively. Compared to PBTFDO, selenium substitution does not affect the planarity of the conjugation backbone, but oxygen substitution compromises this planarity. Contrary to our expectations, the atomic sizes have no significant effect on the degree of conjugation along the oligomer backbone. In fact, the repulsive forces between the lone pair of the heteroatom (i.e., Se, S, O) in the donor unit and the lone pair of the ketone in the neighboring acceptor unit influence the conformation of the conjugation backbone. The lone pair electrons are held more tightly by oxygen than by sulfur or selenium, because the delocalization is not as extensive in furan as it is in thiophene or selenophene. Therefore, when the heteroatom in the donor unit is oxygen, the repulsive forces between the lone pair of this heteroatom and the lone pair of the ketone in the neighboring acceptor unit are maximized leading to large twist angles in the D-A repeat units in PBTFDO(O) and PBTFDO(2O).

Figure 10 shows the relationship of the frontier orbital energy levels between the CP donor acceptor repeat unit (monomer), the isolated donor and acceptor units, as well as hypothetical D-D, and A-A pairs. These results are consistent with our previous findings, which show that the HOMO and LUMO energies of a monomer are determined by the HOMO of the isolated donor unit and the LUMO of the isolated acceptor unit, respectively.

The frontier orbital energy levels of the different length CP oligomers and the periodically continuous configurations are shown in Figure 11. As we increase the number of repeat units from one to four, for a given D-A pairing, the LUMO energy decreases more drastically than the HOMO energy. This is the same trend that was seen in four CPs discussed in section 3.1. However, when the donor unit is varied, the LUMO energy stays approximately constant, whereas the HOMO energy increases in the sequence PBTFDO, PBTFDO(Se), PBTFDO(2Se), PBTFDO(O), PBTFDO(2O). The calculated HOMO/LUMO gaps in these CPs follow the same trend, except that PBTFDO(2Se) and PBTFDO(O) trade places.

By examining the frontier orbitals of the newly designed CPs with the periodically continuous configuration, as shown in Figure 12, we find that the LUMO orbitals are localized on the acceptor units to about the same degree for all the CPs. The HOMO orbitals increasingly withdraw from the acceptor unit and become more localized on the donors, as the sulfur is substituted with one Se, two Se, one O, and finally two O. The increased localization follows the same sequence as the previously calculated HOMO/LUMO gap reductions. This strongly supports our previous finding that the frontier orbital localization strongly correlates with the energy gap decrease for D-A type CPs.

The amount of intramolecular charge transfer in these systems is calculated as the total charge difference associated with the acceptor unit, which is common to all our newly designed CPs. We use the absolute value to be consistent with the charge increase on the donor calculated earlier, i.e., a positive quantity represents a transfer of electrons from donor to acceptor. In Figure 13, we plot the total charge difference for all the CPs vs. the HOMO/LUMO gap, taking the average for all the donor-acceptor positions in the dimers, trimers and tetramers. We see that the greater the intramolecular charge transfer, the smaller the HOMO/LUMO gap, regardless of whether the donating strength of the donor or the withdrawing strength of the acceptor has been manipulated. Moreover, the linear relationship is preserved when adding the newly designed CPs, with an improved correlation coefficient of 0.89. Interestingly, however, the electronegativity values for selenium, sulfur, and oxygen are 2.55, 2.58, and 3.44 respectively.[28] Yet, the electron donating strength of the molecular units these elements reside on behaves inversely proportional to what would be expected based on the tendency of these elements to attract electrons.

## 4 Conclusion

We investigated the factors that control energy level tuning in D-A conjugated polymers using first principles DFT calculations by varying the donor and acceptor units independently of one another. In all cases we observed that upon combining donor and acceptor groups into an isolated D-A repeat unit, the HOMO/LUMO energy levels can be predicted based on HOMO/LUMO energy levels of the isolated donor and acceptor groups, as long as the electron withdrawing or donating powers of the two groups are sufficiently different. As repeat units are assembled into a polymer, the LUMO energy for any given D-A pairing decreases with increasing conjugation length, whereas the HOMO levels remain unaffected by the extended conjugation. Also, the reduction in LUMO energy is more pronounced in acceptor units with higher electronegativity.

Furthermore, the decrease in LUMO energies, is correlated with the extent of the LUMO orbital localization, while the spatial arrangement of the HOMO orbital is exclusively influenced by the chemical nature of the donor group. The HOMO is delocalized for moieties with relatively weak electron donating character, which is the case for the grouping of D-A combinations in which only the acceptor was varied. However, once the electron donating power on the donor is enhanced, as is the case for the grouping subject to elemental substitution within the donor, the HOMO level increases and the HOMO orbitals become more localized.

Finally, increasing the electron withdrawing strength of the acceptor or the donating power of the donor, both result in an increased charge transfer between donor and acceptor, and increased degree of frontier orbital localization, and a decreased HOMO/LUMO energy gap. Reducing the HOMO/LUMO gap in conjugated polymers with alternating D-A acceptor groups will ultimately lead to both HOMO and LUMO orbital localization, but with the LUMO localization preceding that of the HOMO. The pivotal factor for controlling this behavior appears to be electron donating and withdrawing powers of molecular groups, which are, however, not simply derived from the electronegativities of their constituting elements.

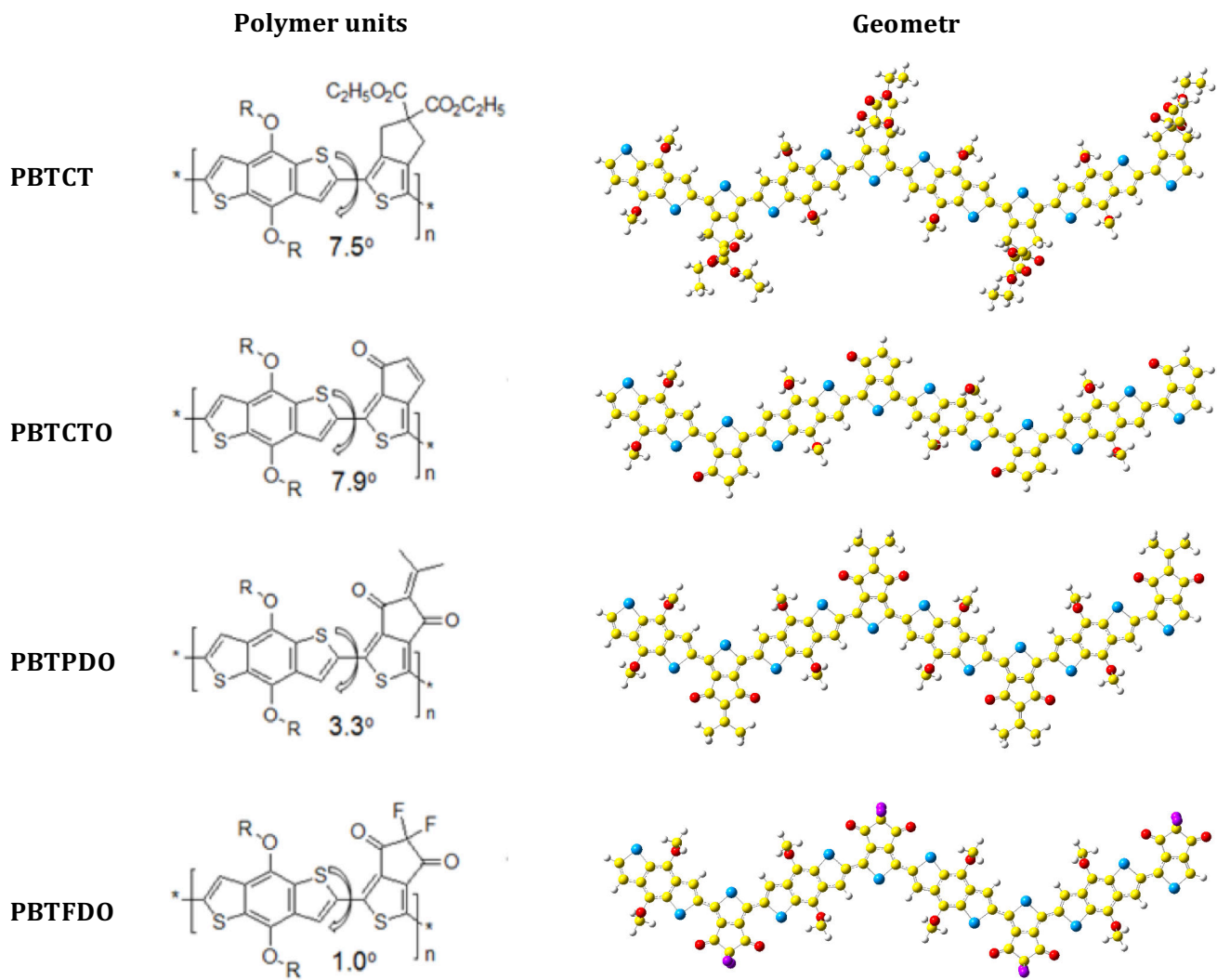

**Figure 1.** Chain conformations of the CPs, obtained from the tetramer conformation calculation under minimized energy state.

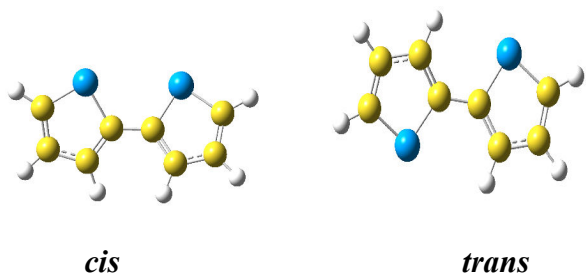

**Figure 2.** The *cis* and *trans* conformations of the bithenyl oligomer.

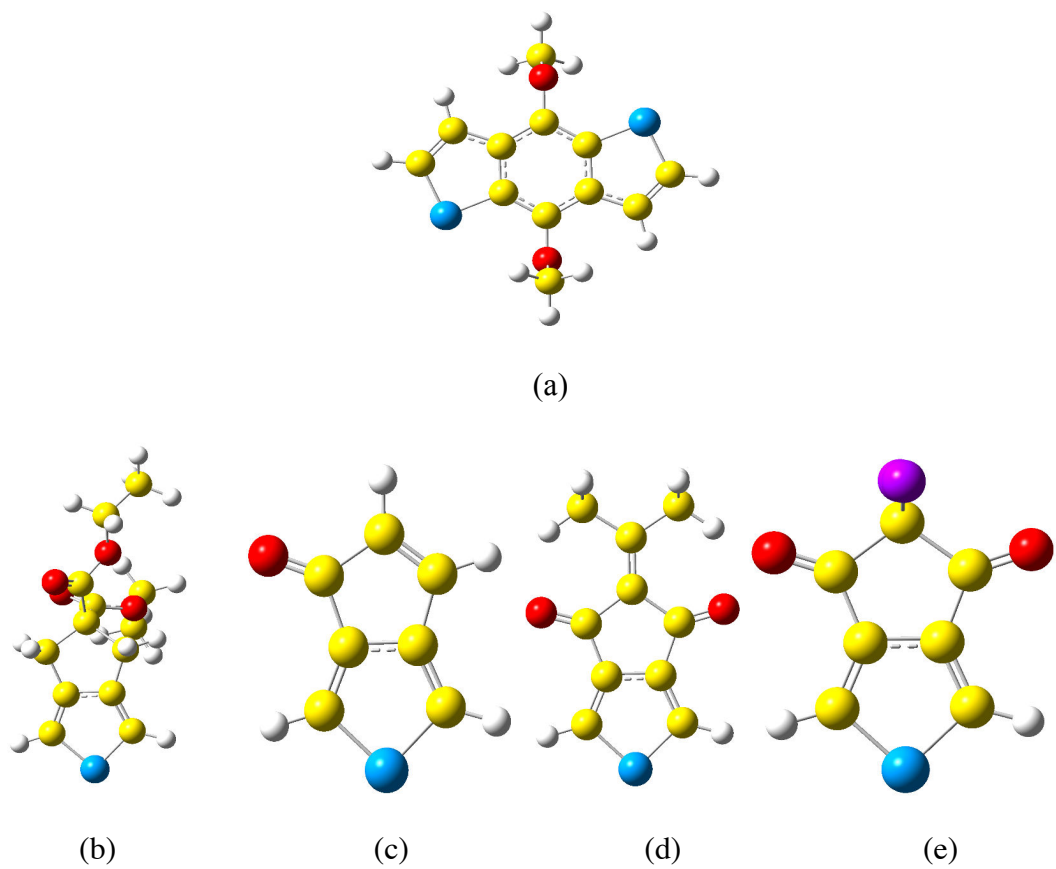

(a)

(b)   (c)   (d)   (e)

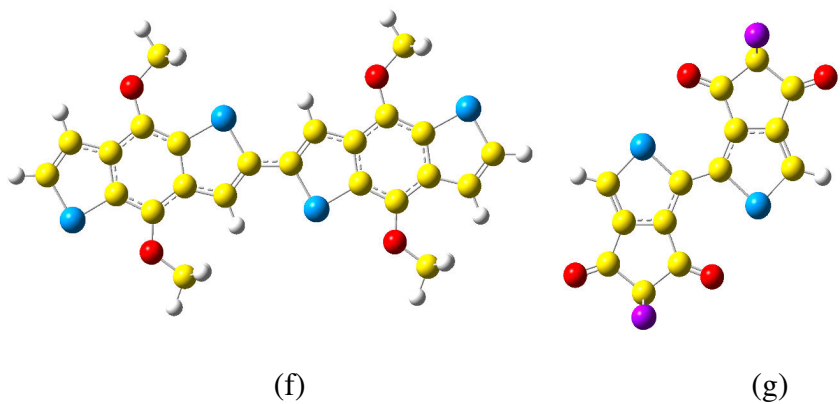

(f) (g)

**Figure 3** (a) hydrogenized donor unit; (b-e) hydrogenized acceptor units for PBTCT, PBTCTO, PBTPDO, PBTFDO from left to right; (f) donor-donor oligomer (g) acceptor-acceptor oligomer based on (e). (yellow) C, (white) H, (blue) S, (red) O, (purple) F.

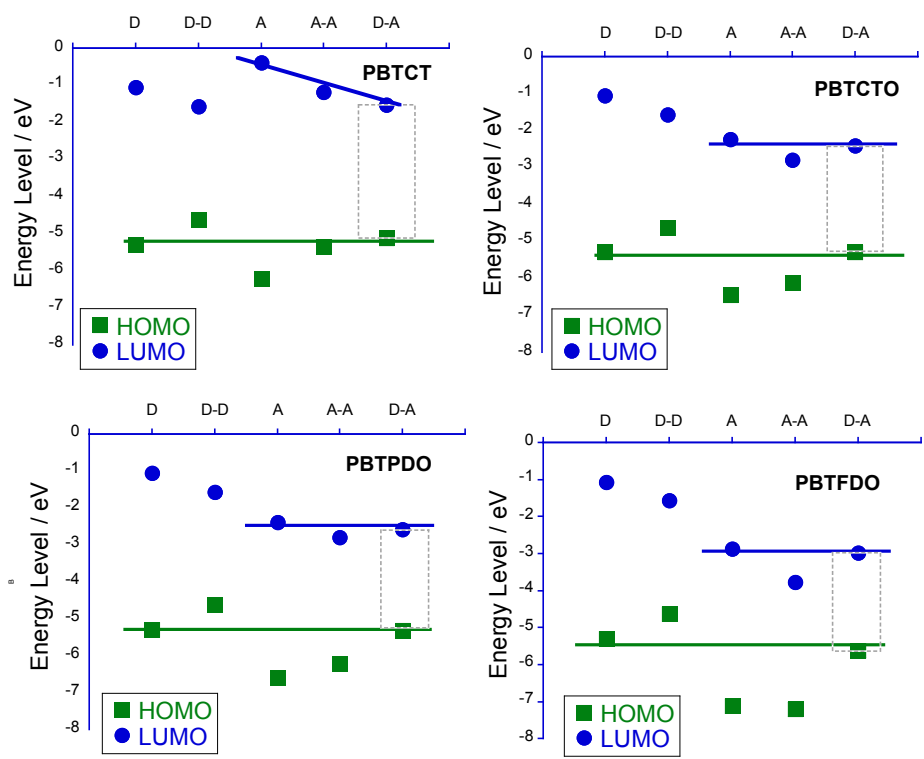

**Figure 4.** The frontier orbital energy levels of donor unit (D), donor-donor unit (D-D), acceptor unit (A), acceptor-acceptor unit (A-A) and CP monomer (D-A). It is found that the HOMO level is determined by that of the donor unit and the LUMO level by that of the acceptor unit in a single D-A repeat unit. The rule applies for all but PBTCT.

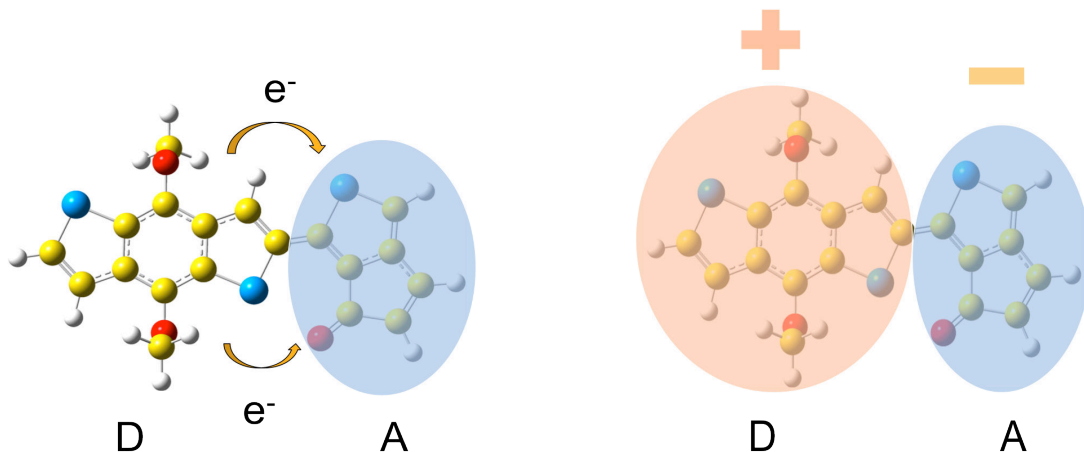

**Figure 5.** An illustration of intramolecular charge transfer for PBTCTO monomer. The charge distribution of donor (acceptor) units in CPs is different from that of corresponding isolated donor (acceptor) unit in Figure 3. The charge redistribution, as a result, can be expressed as the amount of charge transfer from donor unit to neighboring acceptor unit, or vice versa.

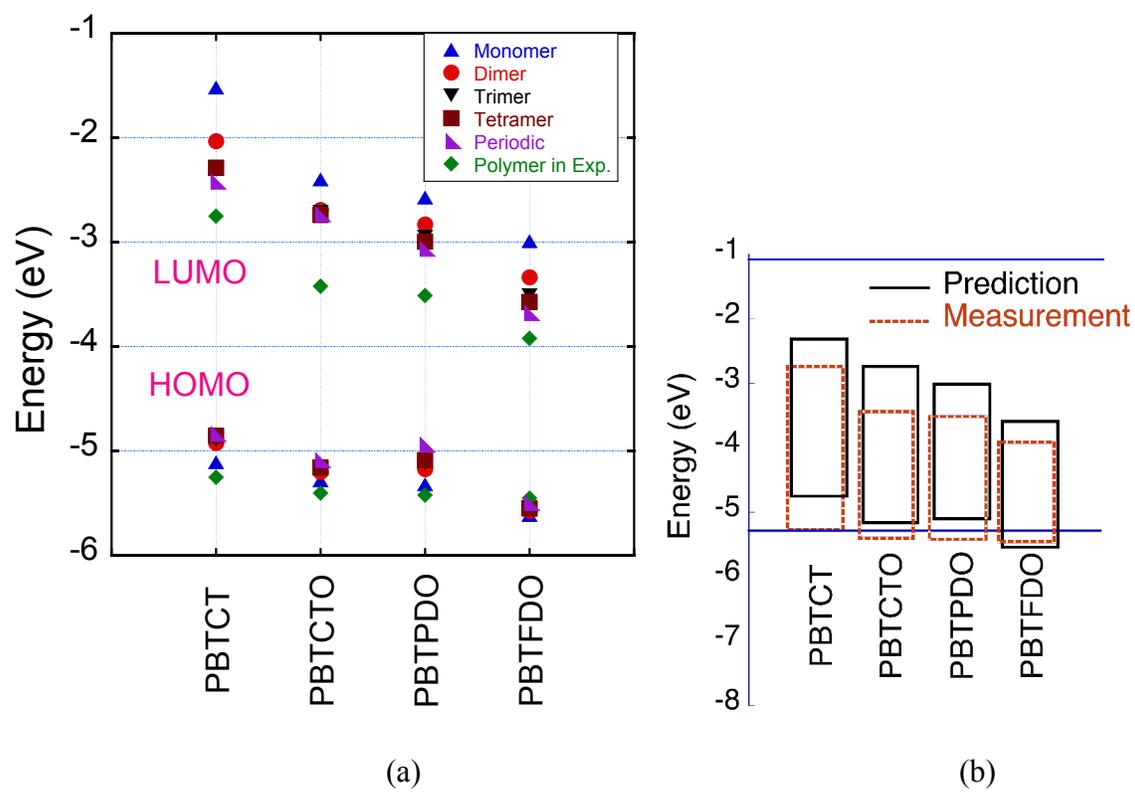

**Figure 6.** (a) Experimental and calculated energy levels, and (b) energy gaps for the CPs. The calculated energy gaps are based on the energy levels of periodic unit cells.

**Table 1** Energy gaps for the CPs (experimental measurement in parenthesis).

| | energy gap (eV) |
|---|---|
| PBTCT | 2.41 (2.50) |
| PBTCTO | 2.36 (1.98) |
| PBTPDO | 1.87 (1.91) |
| PBTFDO | 1.82 (1.53) |

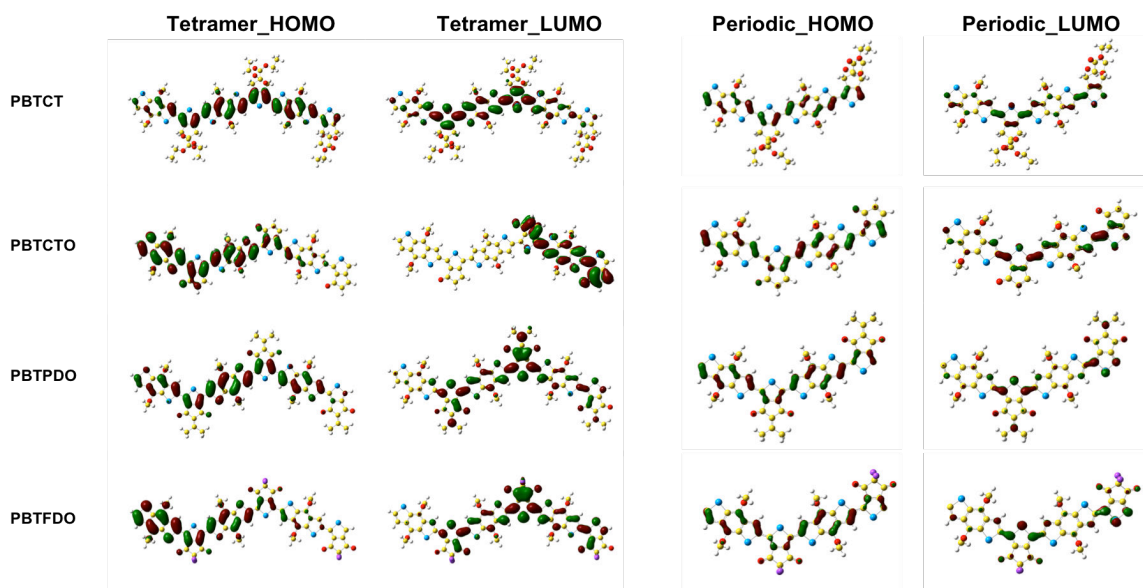

**Figure 7**. The frontier orbitals for CP tetramers and periodic unit cells (a unit cell consists of two D-A repeating units).

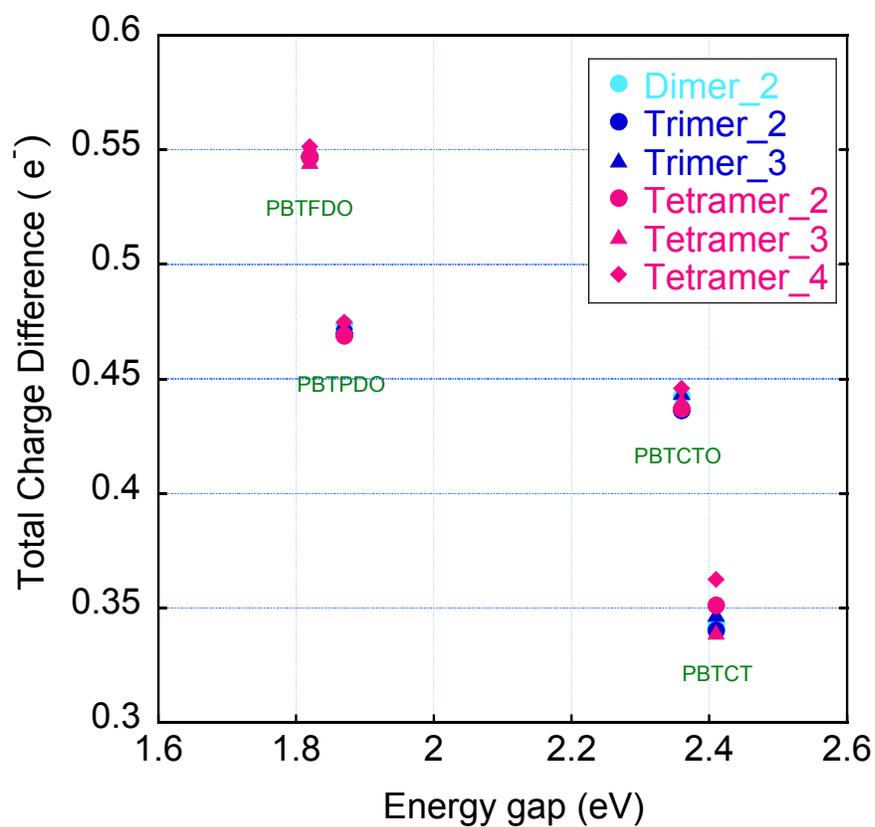

**Figure 8.** The total charge difference of the donor unit in the CPs as a function of energy gap. To eliminate any size effects, only the donor units in the middle of the molecules are considered, e.g. the second donor unit in the dimer is named as Dimer_2 and the third donor unit in the tetramer is named as Tetramer_3.

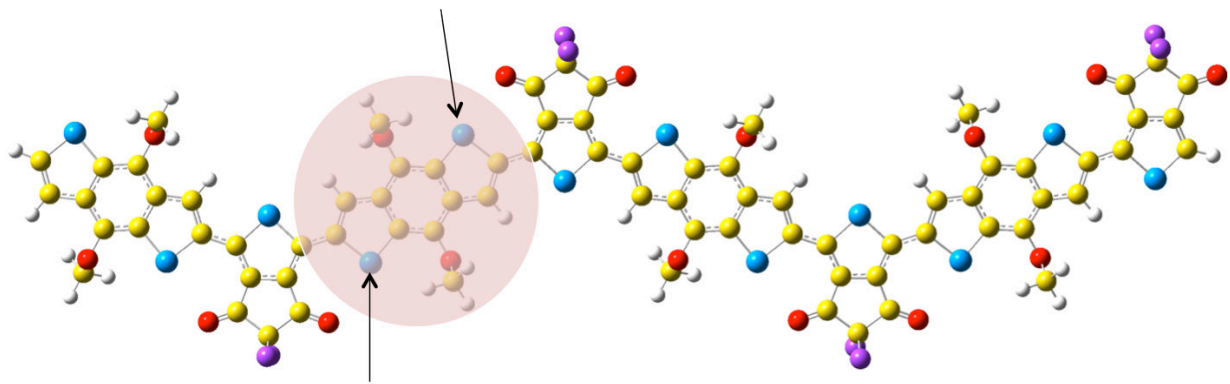

**Figure 9.** The structure of PBTFDO. While maintaining the acceptor unit of PBTFDO, one or two sulfur atoms in the donor unit is replaced by Se or O to obtain a new donor unit, corresponding to four newly designed CPs named as PBTFDO(Se), PBTFDO(2Se), PBTFDO(O), PBTFDO(2O).

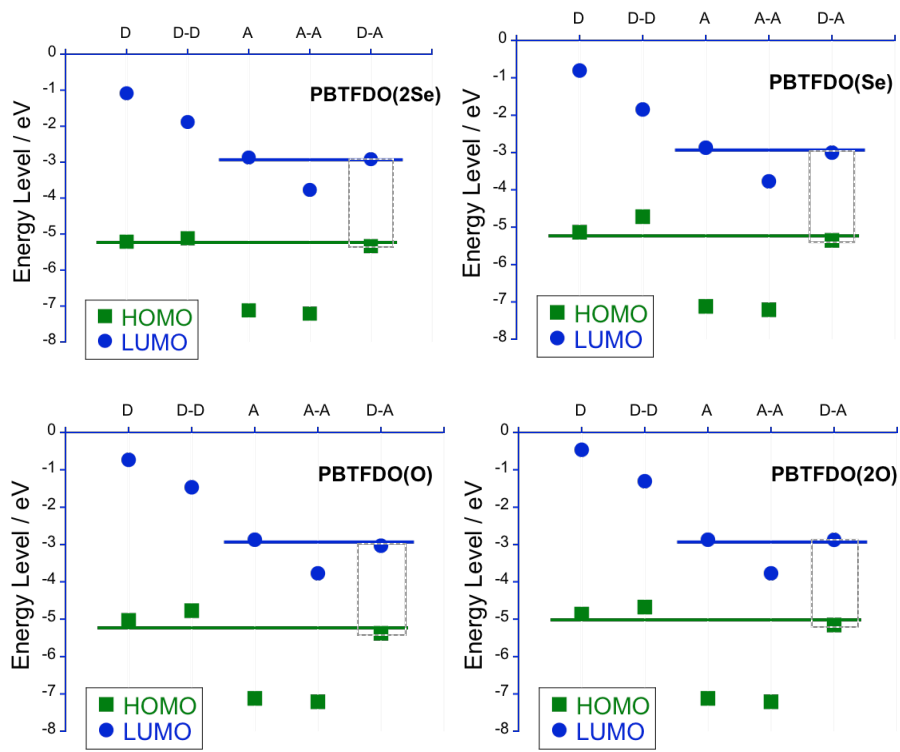

**Figure 10.** The frontier orbital energy levels of donor unit (D), donor-donor unit (D-D), acceptor unit (A), acceptor-acceptor unit (A-A) and CP monomer (D-A) for the newly designed CPs.

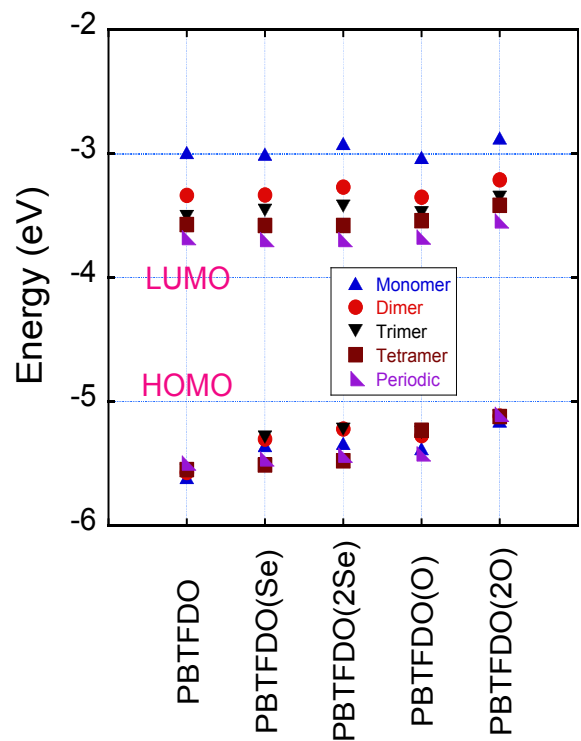

**Figure 11.** Calculated energy levels for the newly designed CPs. PBTFDO is also included for comparison.

**Table 2** Energy gaps for the newly designed CPs. PBTFDO is also included for comparison. The calculated energy gap is based on the energy levels of periodic unit cells

| | energy gap (eV) |
|---|---|
| PBTFDO | 1.82 |
| PBTFDO (Se) | 1.77 |
| PBTFDO (2Se) | 1.74 |
| PBTFDO (O) | 1.75 |
| PBTFDO (2O) | 1.56 |

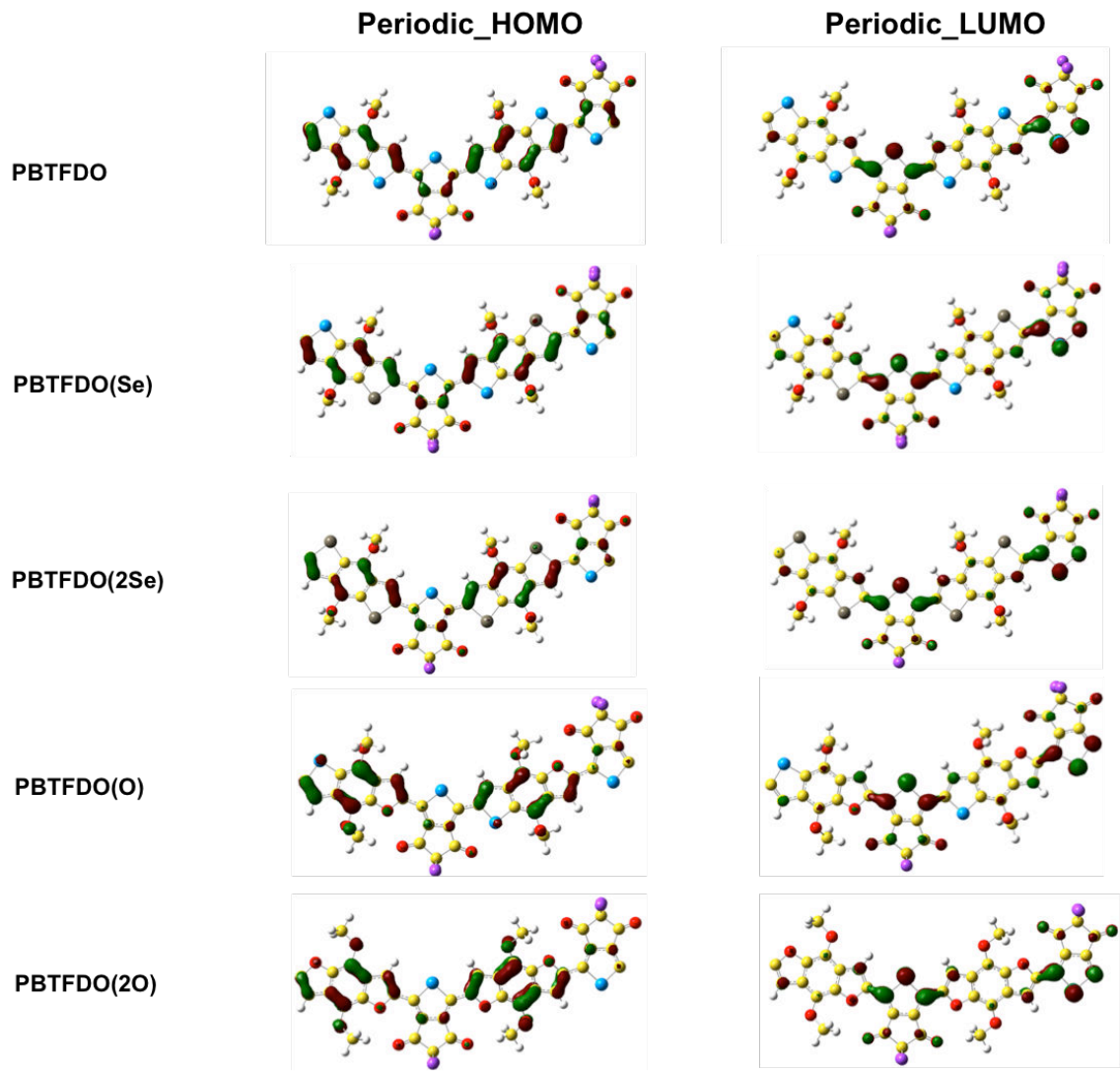

**Figure 12.** The frontier orbitals for the newly designed CP periodic unit cells (a unit cell consists of two D-A repeating units).

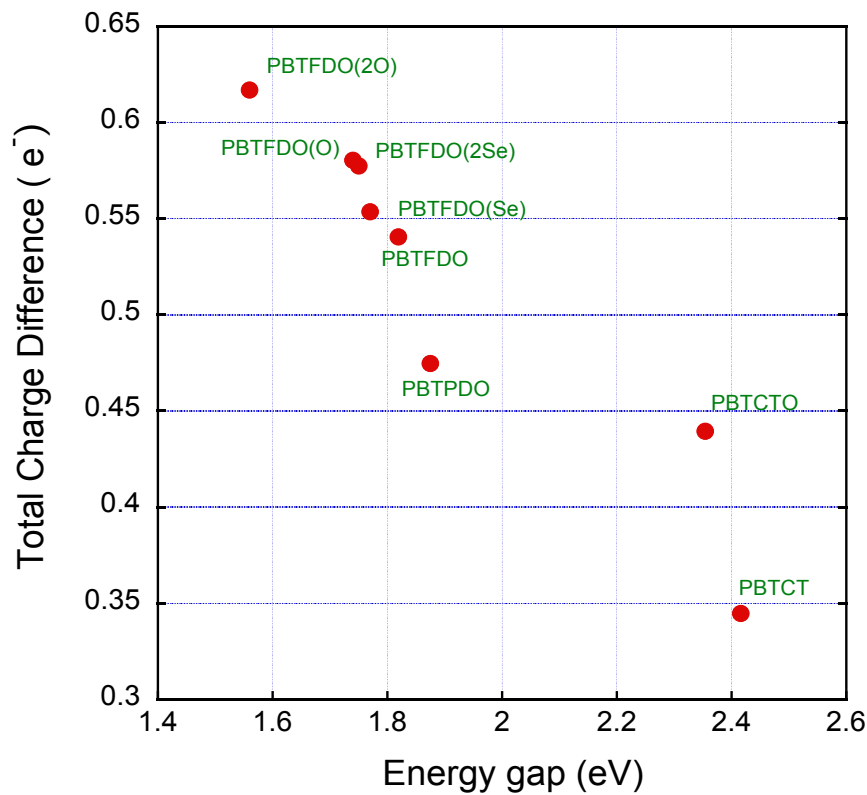

**Figure 13**. The total charge difference as a function of energy gap for all the CPs.